\pdfoutput=1

\documentclass[11pt,twoside,a4paper,cmspaper,final,collab]{cms-tdr}
\def\svnVersion{360573}\def\cmsCernNoTag{CERN-EP/2016-059}\def\cmsCernDate{\today}\def\cmsMessage{Published in Physics Letters B as \href{http://dx.doi.org/10.1016/j.physletb.2016.07.006}{\doi{10.1016/j.physletb.2016.07.006.}}}

\begin{document}\cmsNoteHeader{TOP-15-009}

\hyphenation{had-ron-i-za-tion}
\hyphenation{cal-or-i-me-ter}
\hyphenation{de-vices}
\RCS$Revision: 360343 $
\RCS$HeadURL: svn+ssh://svn.cern.ch/reps/tdr2/papers/TOP-15-009/trunk/TOP-15-009.tex $
\RCS$Id: TOP-15-009.tex 360343 2016-07-27 20:29:17Z linacre $
\newlength\cmsTabSkip\setlength\cmsTabSkip{1.5ex}
\newlength\cmsFigWidth
\ifthenelse{\boolean{cms@external}}{\setlength\cmsFigWidth{0.85\columnwidth}}{\setlength\cmsFigWidth{0.4\textwidth}}
\ifthenelse{\boolean{cms@external}}{\providecommand{\cmsLeft}{top\xspace}}{\providecommand{\cmsLeft}{left\xspace}}
\ifthenelse{\boolean{cms@external}}{\providecommand{\cmsRight}{bottom\xspace}}{\providecommand{\cmsRight}{right\xspace}}
\cmsNoteHeader{TOP-15-009}
\title{\texorpdfstring{Measurements of \ttbar charge asymmetry using dilepton final states in \pp collisions at $\sqrt{s}=8\TeV$}
{Measurements of t t-bar charge asymmetry using dilepton final states in pp collisions at sqrt(s) = 8 TeV}}

\date{\today}

\newcommand{\wjets}{\PW+jets}
\newcommand{\pp}{\ensuremath{\Pp\Pp}\xspace}
\providecommand{\fig}{Fig.}
\providecommand{\tab}{Table}
\providecommand{\reference}{Ref.}
\providecommand{\secn}{Section}

\ifthenelse{\boolean{cms@external}}
{\providecommand{\suppMaterial}{the supplementary material
[URL will be inserted by publisher]\xspace}}
{\providecommand{\suppMaterial}{Appendix~\ref{app:suppMat}\xspace}}

\abstract{The charge asymmetry in \ttbar events is measured using dilepton final states produced in \pp collisions at the LHC at $\sqrt{s}=8\TeV$.
The data sample, collected with the CMS detector, corresponds to an integrated luminosity of 19.5\fbinv. The measurements are performed using events
with two oppositely charged leptons (electrons or muons) and two or more jets, where at least one of the jets is identified as originating from a bottom quark.
The charge asymmetry is measured from differences in kinematic distributions, unfolded to the parton level, of positively and negatively charged top quarks and leptons.
The \ttbar and leptonic inclusive charge asymmetries are found to be $0.011 \pm 0.011\stat \pm 0.007\syst$ and $0.003 \pm 0.006\stat \pm 0.003\syst$, respectively.
These results, as well as charge asymmetry measurements made as a function of
the invariant mass, rapidity, and transverse momentum of the \ttbar system,
are in agreement with predictions of the standard model.
}

\hypersetup{%
pdfauthor={CMS Collaboration},%
pdftitle={Measurements of t t-bar charge asymmetry using dilepton final states in pp collisions at sqrt(s) = 8 TeV},%
pdfsubject={CMS},%
pdfkeywords={CMS, physics, top, charge asymmetry}}

\maketitle

\section{Introduction}
\label{sec:intro}

The exceptionally large mass of the top quark, measured by this experiment as $m_\cPqt = 172.44 \pm 0.48\GeV$~\cite{CMSNewMass},
suggests the top quark could have an important connection to physics beyond the standard model (SM), particularly in the
mechanism of electroweak (EW) symmetry breaking.
Precision measurements of top quark properties have the potential to identify the first hints of new particles, particularly those with stronger couplings to top
quarks than to other fundamental particles. The SM predicts a charge asymmetry in \ttbar production at hadron colliders
through quark-antiquark annihilation. This asymmetry is caused by the interference between the Born and the box diagrams, as well as
between the initial- and final-state radiation diagrams, and is predicted by quantum chromodynamics (QCD) calculations at next-to-leading order (NLO)~\cite{Kuhn:1998jr,Kuhn:1998kw}.
Early measurements of this asymmetry by the CDF~\cite{:2011kc} and D0~\cite{PhysRevD.84.112005} collaborations
exceeded the NLO predictions~\cite{Kuhn:1998jr,Kuhn:1998kw} by about
two standard deviations, and the discrepancy was more pronounced in the CDF events with large \ttbar invariant mass ($M_{\ttbar}>450\GeV$).
These results have led to considerations that the anomalous asymmetry might be generated by
tree-level exchanges of new particles or by interference effects from new physics at higher mass scales, not directly observable at the LHC~\cite{RevModPhys.87.421}.
Recent developments in experimental
techniques~\cite{ref:CDFafb2013,PhysRevD.90.072011} and theoretical predictions such as the inclusion of EW~\cite{Bernreuther:2010ny,PhysRevD.84.093003,Kuhn:2011ri,ref:BernChargeAsym}
and next-to-next-to-leading-order (NNLO) QCD~\cite{PhysRevLett.115.052001,Czakon2016} corrections
have largely resolved the disagreement between theory and the Tevatron measurements.
Nonetheless, the charge asymmetry remains an important probe of new physics.

At the Tevatron, colliding valence quarks from the proton and antiproton beams result in asymmetric rapidity ($y$) distributions of top quarks and antiquarks.
The proton-proton (\pp) initial state at the LHC is expected to produce top quark and antiquark rapidity distributions that are symmetric about $y = 0$.
However, since the quarks in the initial state can be from valence, while the antiquarks are
from the sea, the larger average momentum-fraction of quarks leads to an excess of top quarks produced in the forward directions. The rapidity distribution of
top quarks in the SM is therefore broader than that of the more centrally produced top antiquarks, meaning $\Delta\abs{y_{\cPqt}} = \abs{y_{\cPqt}} -
\abs{y_{\cPaqt}}$ is a suitable observable to measure the \ttbar charge asymmetry,
defined in terms of event yields $N$ as
\begin{equation*}
A_{\mathrm{C}} = \frac{N(\Delta\abs{y_{\cPqt}} > 0)-N(\Delta\abs{y_{\cPqt}} < 0)}{N(\Delta\abs{y_{\cPqt}} > 0)+N(\Delta\abs{y_{\cPqt}} < 0)}.
\end{equation*}
While the measurement of $A_{\mathrm{C}}$ relies on the reconstruction
of the top quark and antiquark directions, an advantage of the dilepton final state is that one can alternatively measure the
leptonic charge asymmetry defined using only the lepton pseudorapidities~\cite{JINST} $\eta_{\ell^\pm}$ as
\begin{equation*}
A^{\text{lep}}_{\mathrm{C}} = \frac{N (\Delta\abs{\eta_{\ell}}>0)-N (\Delta\abs{\eta_{\ell}}<0)}{N (\Delta\abs{\eta_{\ell}}>0)+N (\Delta\abs{\eta_{\ell}}<0)},
\end{equation*}
where $\Delta\abs{\eta_{\ell}} = \abs{\eta_{\ell^+}} - \abs{\eta_{\ell^-}}$.
This observable is useful because it is free of the ambiguities associated with the top quark reconstruction, and because
the correlation between the direction of a top quark and its decay products transmits an asymmetry in the parent top quarks to the daughter leptons.
Furthermore, its dependence on the top quark polarization implies that it is not fully correlated with $A_C$ and provides complementary information~\cite{ref:Krohn}.
Previous ATLAS and CMS measurements of $A_{\mathrm{C}}$ using data from \pp collisions at
 $\sqrt{s}=7\TeV$~\cite{semileptonicCMS,Aad:2013cea} and $8\TeV$~\cite{1507.03119,1508.03862,1509.02358,1512.06092}, and of $A^{\text{lep}}_{\mathrm{C}}$ using the $7\TeV$ data samples~\cite{Chatrchyan:2014yta,Aad:2015jfa}, are consistent with the SM predictions.

In this Letter, measurements are presented of $A_{\mathrm{C}}$ and $A^{\text{lep}}_{\mathrm{C}}$ from \ttbar\ events in the dilepton final states,
using CMS data from \pp collisions at
$\sqrt{s}=8\TeV$
 corresponding to an integrated luminosity of 19.5\fbinv.
The analysis strategy is similar to that presented in \reference~\cite{Chatrchyan:2014yta} with many improvements, most importantly
in the unfolding technique.
This allows for full differential measurements of $A_{\mathrm{C}}$ and $A^{\text{lep}}_{\mathrm{C}}$, which are made as a function of $M_{\ttbar}$ as well as the absolute rapidity and the transverse momentum of the \ttbar\ system in the laboratory frame ($\abs{y_{\ttbar}}$ and $\pt^{\ttbar}$).
Furthermore, the larger data sample used here as well as improvements made in the resolution of the top quark reconstruction lead to better statistical precision.

\section{The CMS detector}

The central feature of the CMS apparatus is a superconducting solenoid of 6\unit{m} internal diameter, providing a magnetic field of 3.8\unit{T}. Within the solenoid volume are a silicon pixel and strip tracker, a lead tungstate crystal electromagnetic calorimeter, and a brass and scintillator hadron calorimeter, each composed of a barrel and two endcap sections. Forward calorimeters extend the pseudorapidity coverage provided by the barrel and endcap detectors.
Gas-ionization detectors embedded in the steel flux-return yoke outside the solenoid provide additional measurements of muons.
A more detailed description of the CMS detector, together with a definition of the coordinate system used and the relevant kinematic variables, can be found in \reference~\cite{JINST}.

\section{Event selection and reconstruction}
\label{sec:eventsel}

The event selection for this analysis is identical to that used in \reference~\cite{SpinCorr8TeV} and is only briefly described in this section.
The particle-flow (PF) method~\cite{CMS-PAS-PFT-09-001,CMS-PAS-PFT-10-001} is used to reconstruct final-state particles.
Events are required to have exactly two isolated~\cite{SpinCorr8TeV} leptons (electrons~\cite{Khachatryan:2015hwa} or muons~\cite{MUOART}) of opposite electric charge, with $\pt > 20\GeV$ and
$\abs{\eta} < 2.4$.
The dilepton pair invariant mass $M_{\ell\ell}$ is required to be above $20\GeV$.
For same-flavor leptons, $M_{\ell\ell}$ must also not be within $15\GeV$ of the \cPZ~boson mass
to suppress the Drell--Yan (\ensuremath{\cPZ/\Pgg^\star}+jets) background.

The anti-\kt clustering algorithm~\cite{antikt} with a distance parameter of 0.5
is used to form jets from the PF objects.
The contribution to the jet energy from additional interactions in the same bunch crossing (pileup) is estimated for each event using the
jet area method~\cite{cacciari-2008-659}, and is subtracted from the overall jet \pt.
At least two jets with $\pt > 30\GeV$ and  $\abs{\eta} < 2.4$ are required in each event.
At least one of these jets must be consistent with containing
the decay of a heavy-flavor hadron, as identified using the medium
operating point of the combined secondary vertex (CSV)
\cPqb~tagging algorithm~\cite{ref:btag}. We refer to such jets as \cPqb-tagged jets.

The missing transverse momentum vector \ptvecmiss is defined as the negative vector sum of the
\pt of all PF objects over the full calorimeter coverage ($\abs{\eta} < 5$).
Its magnitude is referred to as \MET.
The calibrations that are applied to the energy measurements of jets are propagated to a correction of \ptvecmiss.
The \MET\ value is required to exceed $40\GeV$ in events with same-flavor leptons in order to further suppress the Drell--Yan background.
There is no \MET\ requirement for $\Pe^{\pm}\Pgm^{\mp}$ events.

The inclusive measurement of $A_{\mathrm{C}}$ and all differential measurements presented here require reconstruction of the \ttbar\ system.
Each signal event has two neutrinos, and there is also a twofold ambiguity in combining the \cPqb~jets with the leptons.
In 62\% of the events passing the event selection requirements, only one of the selected jets is \cPqb~tagged. In those events the untagged jet with the highest ranking by the CSV algorithm
is assumed to be the second \cPqb~jet. Solutions for the neutrino momenta are found analytically assuming $m_\cPqt=172.5\GeV$.
Each event can have up to 8 possible solutions, and the one with the maximum weight obtained using the matrix weighting technique~\cite{ref:1105.5661} is chosen
as the most probable.
For events with no physical solution,
we attempt to find a solution for the sum of neutrino \pt as close as possible to the measured \ptvecmiss~\cite{Dalitz1992225,Betchart:2013nba}.
Nonetheless, no solution is found for approximately 16\% of the events, both in data and simulation. Events with no solutions are used only
in the inclusive measurement of $A^{\text{lep}}_{\mathrm{C}}$, although the results do not significantly change if those events are excluded.
The signs of $\Delta\abs{y_{\cPqt}}$ and $\Delta\abs{\eta_{\ell}}$ are correctly reconstructed in $74.9\%$ and $99.5\%$ of selected simulated \ttbar\ events, respectively.

\section{Event samples and background estimation}
\label{Sec:BkgEst}

The simulated \ttbar\ events used in this analysis are generated using the \MCATNLO3.41~\cite{mc@nlo,MCatNLO2} Monte Carlo (MC) event generator,
with $m_\cPqt=172.5\GeV$ and the CTEQ6M parton distribution functions (PDFs)~\cite{Pumplin:2002vw}.
The subsequent parton showering and fragmentation are done using \HERWIG6.520~\cite{herwig6}.
Simulations with different values of $m_\cPqt$ and the renormalization and factorization scales ($\mu_\mathrm{R}$ and $\mu_\mathrm{F}$) are used to evaluate the associated
systematic uncertainties.
Events with dileptonic \ttbar\ decays, including tau leptons that decay leptonically, are defined as signal, while all other \ttbar\ decay modes are treated as background.
Background events from the \wjets, Drell--Yan, diboson ($\PW\PW$, $\PW\cPZ$, and $\cPZ\cPZ$),  triboson, and $\ttbar$+boson
processes are generated with \MADGRAPH~5.1.3.30~\cite{Alwall:2011uj,MADGRAPHNEW}, while single top quark
events are generated using \POWHEG1.0~\cite{Nason:2004rx,Frixione:2007vw,Alioli:2009je,Alioli:2010xd,Re:2010bp}.
The parton showering and fragmentation are performed using \PYTHIA6.4.22~\cite{Pythia},
which is also used for an alternative \ttbar\ event sample generated using \POWHEG.
Cross sections calculated to NLO or NNLO are used to normalize the background
samples~\cite{xsec_MCFM,Campbell:2011bn,MCatNLO3,xsec_ttbarW,xsec_ttbarZ,xsec_WZ,Kidonakis:2011wy,Kidonakis:2010tc,Kidonakis:2010ux}.

For all MC generated events, pileup is simulated with \PYTHIA\ and superimposed on the hard collisions
using a pileup multiplicity distribution that reflects the luminosity profile of the analyzed data.
The CMS detector response is simulated using a \GEANTfour-based model~\cite{Geant}, and
the events are reconstructed and analyzed with the same software used to process the data.
The measured trigger efficiencies are used to weight the simulated events to account for the trigger requirement, while
the lepton selection efficiencies (reconstruction, identification, and isolation) are consistent between data and
simulation~\cite{Chatrchyan:2013xna,SpinCorr8TeV}. The differences between \cPqb~tagging efficiencies measured in data and simulation~\cite{ref:btag} are
accounted for using correction factors.

The total contribution from background events to the data sample is expected to be 9\%, of which about half comes from single top quark production in association with a \PW~boson ($\PQt\PW$), with dileptonic decays.
Several control regions (CRs) in data are used to validate
the background estimates from simulation for $\PQt\PW$ and $\Z/\gamma^{*}$+jets production and for
events with incorrectly identified leptons.
The CRs are selected to have similar kinematic properties to the signal region, but
with one or two requirements inverted, thus enriching them in different background contributions~\cite{SpinCorr8TeV}.
Agreement between data and simulation is observed in the $\PQt\PW$ CR, and we assign a
 25\% uncertainty in the $\PQt\PW$ cross section based on the recent CMS measurement of $23.4 \pm 5.4$\unit{pb}~\cite{Chatrchyan:2014tua}.
The other CRs are used to derive scale factors (SFs) to multiply the simulated event yields for the corresponding background process, with systematic uncertainties estimated from the envelope of variation in the SF value using the three
dilepton flavor combinations and various alternative CRs.

Other processes, including
\ttbar\ production in association with a boson as well as diboson and triboson production,
contribute less than 20\% of the total background and
 are estimated from simulation alone. Recent CMS measurements~\cite{ref:WWWZ,Chatrchyan:2014aqa,Khachatryan:2015sha}
 indicate agreement between the predicted and measured cross sections for these processes,
and their small yields permit the choice of a conservative systematic uncertainty of 50\% with negligible effect on the analysis precision.

A comparison of the observed and predicted distributions of $\Delta\abs{y_{\cPqt}}$ and $\Delta\abs{\eta_{\ell}}$ can be found in \suppMaterial.

\section{Unfolding the distributions}
\label{sec:unfolding}
The measured distributions are distorted, relative to the true underlying distributions,
by the acceptance of the detector, the efficiency of the trigger and event selection, and the finite
resolution of the reconstructed kinematic quantities.
After subtraction of the predicted background, we correct the measured distributions for these effects using an unfolding procedure that
estimates the corresponding parton-level distributions.
In the context of theoretical calculations and parton shower event generators, the parton-level top quark is defined before it decays and its kinematic properties include the effects of recoil
from initial- and final-state radiation in the rest of the event and from final-state radiation from the top quark itself. The parton-level charged lepton, produced from the
decay of the intermediate \PW~boson, is defined before the lepton decays or radiates any photons.

We use six bins of varying width in the $\Delta\abs{y_{\cPqt}}$ parton-level distribution that are well matched to the reconstruction resolution and contain approximately equal numbers of events.
The $\Delta\abs{\eta_{\ell}}$ distribution depends only on lepton measurements,
and the better resolution allows us to use 12 bins. For the
reconstruction-level distributions, we use twice as many bins as those used for the parton-level distributions.
The unfolding is performed using the \textsc{TUnfold} package~\cite{1748-0221-7-10-T10003},
using regularization based on the curvature of the simulated signal distribution to suppress statistical fluctuations in the high frequency components of the unfolded distribution.
The regularization strength is optimized by minimizing the average global correlation coefficient in the unfolded distribution;
the resulting regularization is relatively weak, contributing at the level of 5\% to the total $\chi^2$ minimized by the algorithm.
An analogous unfolding procedure is used to measure $A_{\mathrm{C}}$ and $A^{\text{lep}}_{\mathrm{C}}$ differentially, after introducing a further three bins in each of the \ttbar\ system kinematic variables $M_{\ttbar}$, $\abs{y_{\ttbar}}$, and $\pt^{\ttbar}$.

\section{Systematic uncertainties}	
\label{sec:systematics}

Most of the systematic uncertainties
concern detector performance and the modeling of the signal and background processes and
are estimated from the change in the measurement when varying the simulated event samples used for the unfolding.
The uncertainty from the jet energy scale corrections
is estimated by varying the jet energies within their uncertainties~\cite{Chatrchyan:2011ds} and propagating this to the \ptvecmiss.
Similarly, the jet energy resolution is varied by 2--5\%, depending on the
$\eta$ of the jet~\cite{Chatrchyan:2011ds}, and
the electron energy scale is varied
by ${\pm}0.6\%$ (${\pm}1.5\%$) for barrel (endcap) electrons, as estimated from comparisons between
measured and simulated \cPZ~boson events~\cite{Khachatryan:2015hwa}.
The uncertainty in muon energies is negligible.
The uncertainty in the background subtraction is obtained by varying the normalization of each
background component by the uncertainties described in \secn~\ref{Sec:BkgEst}.

Many of the signal modeling and simulation uncertainties are evaluated by using weights to vary the \MCATNLO \ttbar\ sample:
the simulated pileup multiplicity distribution is changed within its uncertainty;
the correction factors between data and simulation for the \cPqb~tagging efficiency~\cite{ref:btag}, trigger efficiency, and lepton selection
efficiency are shifted up and down by their uncertainties;
and the PDFs are varied using the PDF4LHC procedure~\cite{Alekhin:2011sk,pdf4lhcInterim}.
Previous CMS studies~\cite{toppT,Khachatryan:2015oqa} have shown that the \pt\ distribution of the top quark in data is softer than in the NLO
simulation of \ttbar\ production.
Since the origin of the discrepancy is not fully understood, the change in the measurement when reweighting
the \MCATNLO \ttbar\ sample to match the top quark \pt\ spectrum in data is taken as a systematic uncertainty associated with signal modeling.
Further signal modeling uncertainties are evaluated using the dedicated \ttbar\ samples:
$\mu_\mathrm{R}$ and $\mu_\mathrm{F}$ are simultaneously varied up and down by a factor of 2,
$m_\cPqt$ is varied by $\pm1\GeV$,
and
the \ttbar\ sample generated with \POWHEG and \PYTHIA is used
to measure the uncertainty in hadronization modeling from the difference between the \HERWIG and \PYTHIA descriptions.
The systematic uncertainty estimates evaluated using dedicated \ttbar\ samples have a significant statistical uncertainty governed by the number of events in the simulated samples.
To avoid underestimation of these uncertainties, the maximum of the estimated systematic uncertainty and the statistical uncertainty in that estimate is taken as the final systematic uncertainty.

\begin{table*}[htb]															
\centering	
\topcaption{\label{tab:asyms_sys} Systematic uncertainties in the inclusive values of the charge asymmetries obtained from the unfolded distributions. Uncertainties of less than $0.0005$ are marked by a dash (--). }		
\begin{tabular}{l r r}															
\hline
Charge asymmetry variable		&	\multicolumn{1}{c}{$A_{\mathrm{C}}$}		&	\multicolumn{1}{c}{$A^{\text{lep}}_{\mathrm{C}}$} \\	
\hline																												
\multicolumn{1}{l}{\textit{Experimental systematic uncertainties}}	\\[\cmsTabSkip]																												
Jet energy scale                  & 0.001   & \multicolumn{1}{c}{--}   \\
Jet energy resolution             & 0.002   & \multicolumn{1}{c}{--}   \\
Lepton energy scale               & 0.001   & \multicolumn{1}{c}{--}   \\
Background                        & 0.001   & 0.001   \\
Pileup                            & \multicolumn{1}{c}{--}& \multicolumn{1}{c}{--}   \\
\cPqb~tagging efficiency          & 0.001   & \multicolumn{1}{c}{--}   \\
Lepton selection                  & \multicolumn{1}{c}{--}& \multicolumn{1}{c}{--}   \\
\hline\\[-2.3ex]
\multicolumn{1}{l}{\textit{$t\bar{t}$ modeling uncertainties}}	\\[\cmsTabSkip]																													
Parton distribution functions     & 0.001   & 0.001   \\
Top quark \pt                     & 0.001   & \multicolumn{1}{c}{--}   \\
Renormalization and factorization scales         & 0.003   & 0.002   \\
Top quark mass                    & 0.001   & 0.001   \\
Hadronization                     & 0.003   & \multicolumn{1}{c}{--}   \\
\hline
Unfolding (simulation statistical) & 0.005   & 0.002   \\
Unfolding (regularization)        & \multicolumn{1}{c}{--}   & \multicolumn{1}{c}{--}   \\
\hline																														
Total systematic uncertainty      & 0.007   & 0.003   \\
\hline															
\end{tabular}																														
\end{table*}	

The uncertainty in the unfolding procedure is dominated by the statistical uncertainty arising from the limited number of events in the \MCATNLO \ttbar\ sample. The uncertainty from the regularization is found to be small in comparison.
The systematic uncertainties in the inclusive charge asymmetry values obtained from the unfolded distributions are summarized in \tab~\ref{tab:asyms_sys}.
The individual terms are added in quadrature to estimate the total systematic uncertainties.
For both $A_{\mathrm{C}}$ and $A^{\text{lep}}_{\mathrm{C}}$, the dominant systematic uncertainty arises from the limited number of simulated events used for the unfolding.

\section{Results}
\label{sec:results}

\begin{figure}[htbp]
\centering
\includegraphics[width=0.49\textwidth]{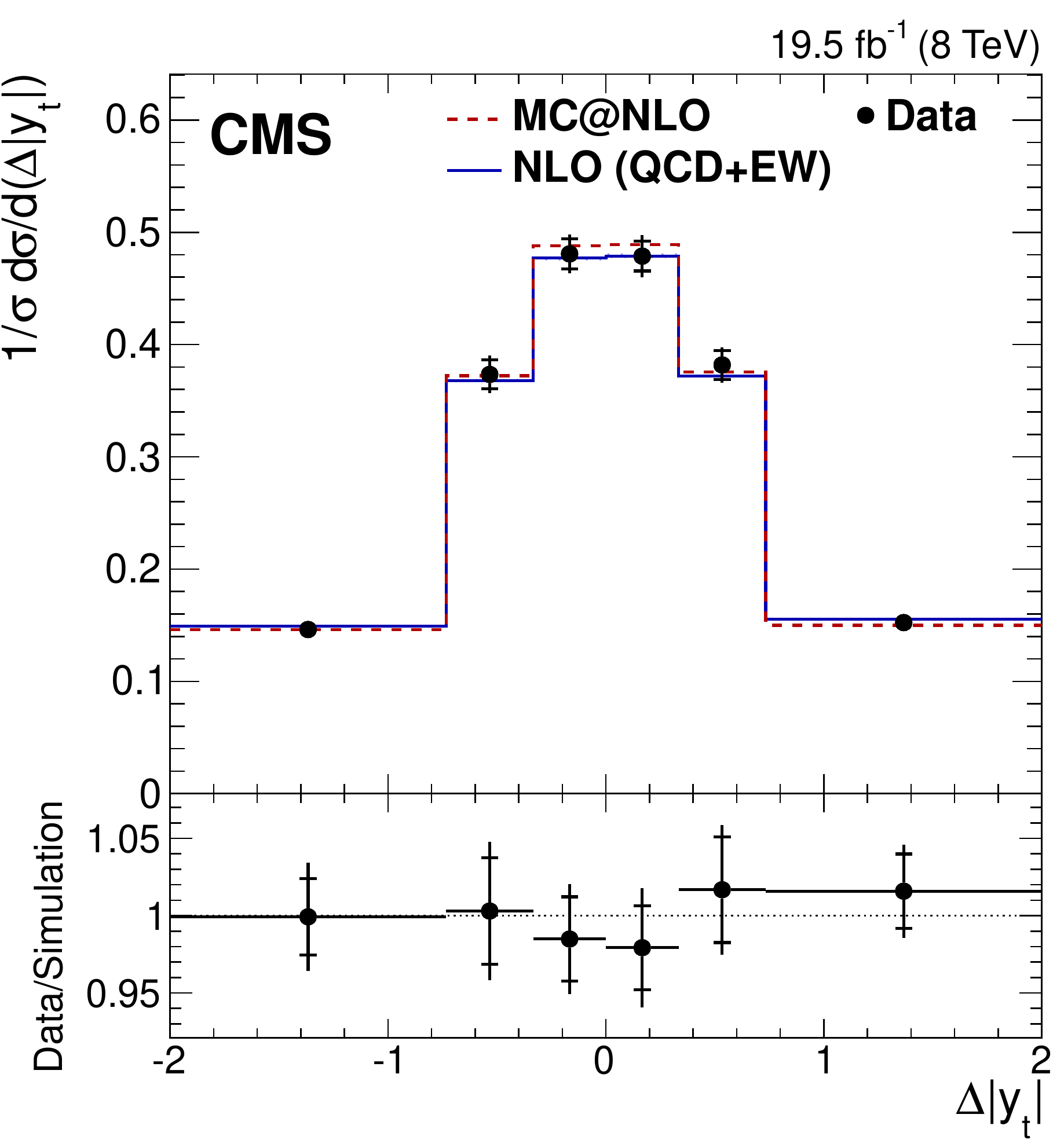}
\includegraphics[width=0.49\textwidth]{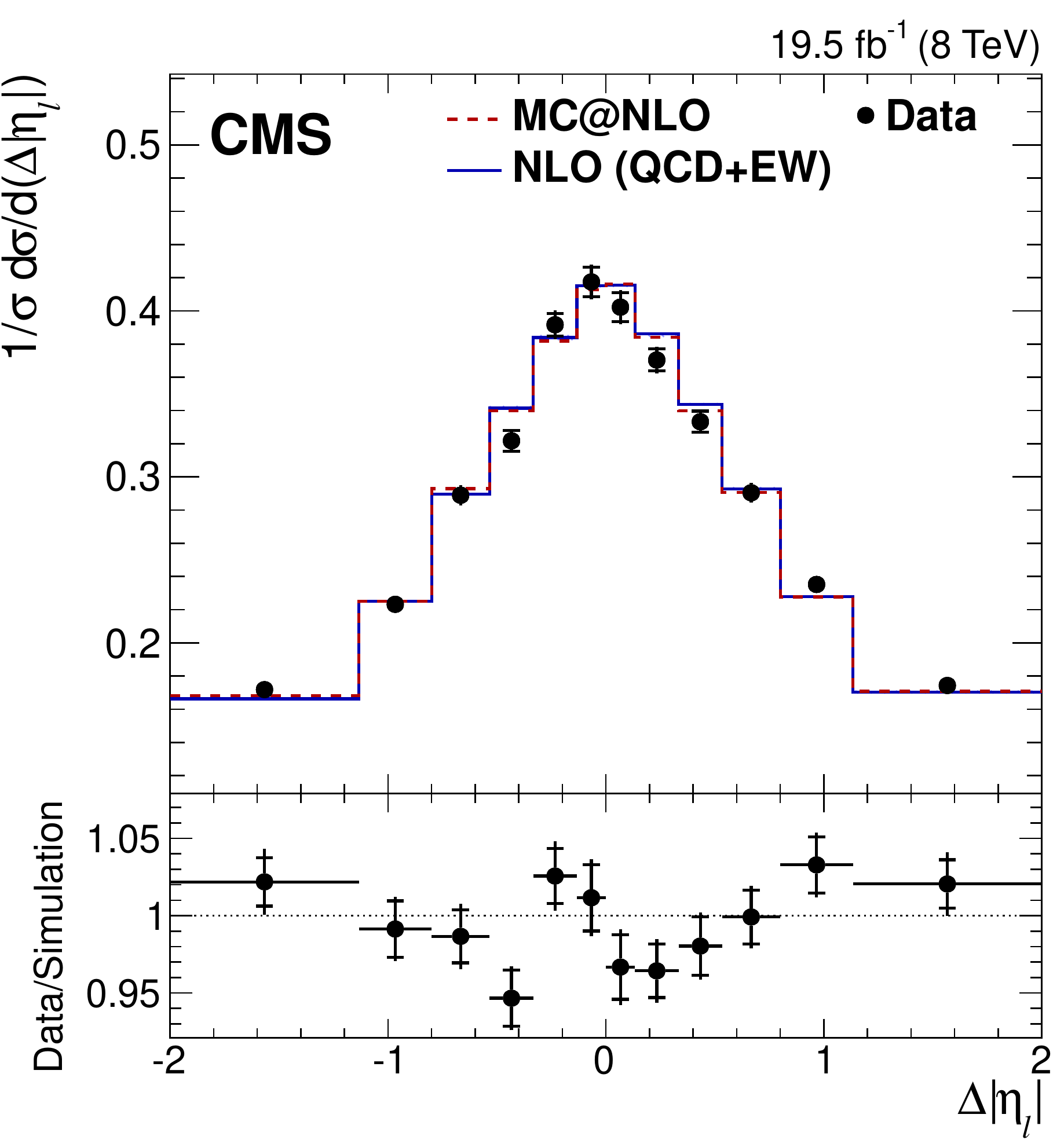}
\caption{\label{fig:ResultsUnfolded}\protect
Background-subtracted and unfolded distributions of $\Delta\abs{y_{\cPqt}}$ (\cmsLeft) and $\Delta\abs{\eta_{\ell}}$ (\cmsRight) from data (points), normalized to unit area.
Parton-level predictions from the \MCATNLO\ simulation and calculations at NLO (QCD+EW)~\cite{ref:BernChargeAsym} are shown by dashed and solid histograms, respectively.
The ratio of the measured bin values to the \MCATNLO\ prediction is shown in the bottom panel.
The vertical bars show the total uncertainty, the statistical component of which is marked by a horizontal tick.
The first and last bins of each plot include underflow and overflow events, respectively.
}
\end{figure}

\begin{table*}[htbp]
\centering
\topcaption{\label{tab:ResultsUnfolded}
The inclusive charge asymmetry measurements obtained from the unfolded distributions and the parton-level predictions from
the \MCATNLO\ simulation and calculations at NLO (QCD+EW)~\cite{ref:BernChargeAsym}.
For the data, the first uncertainty is statistical and the second is systematic.
The uncertainties in the \MCATNLO\ results are statistical and the uncertainties in the NLO calculations come from varying together $\mu_\mathrm{R}$ and $\mu_\mathrm{F}$ up and down by a factor of two.
}
\begin{tabular}{lccc}
\hline
Variable &  Data  &  \MCATNLO & NLO (QCD+EW) \\ \hline
$A_{\mathrm{C}}$		&  $0.011 \pm 0.011 \pm 0.007$  &  $ 0.006 \pm 0.001$   &  $0.0111 \pm 0.0004$ \\
$A^{\text{lep}}_{\mathrm{C}}$	&  $0.003 \pm 0.006 \pm 0.003$  &  $ 0.004 \pm 0.001$   &  $0.0064 \pm 0.0003$ \\
\hline
\end{tabular}
\end{table*}

\begin{figure*}[hbtp]
\centering
\includegraphics[width=0.40\textwidth]{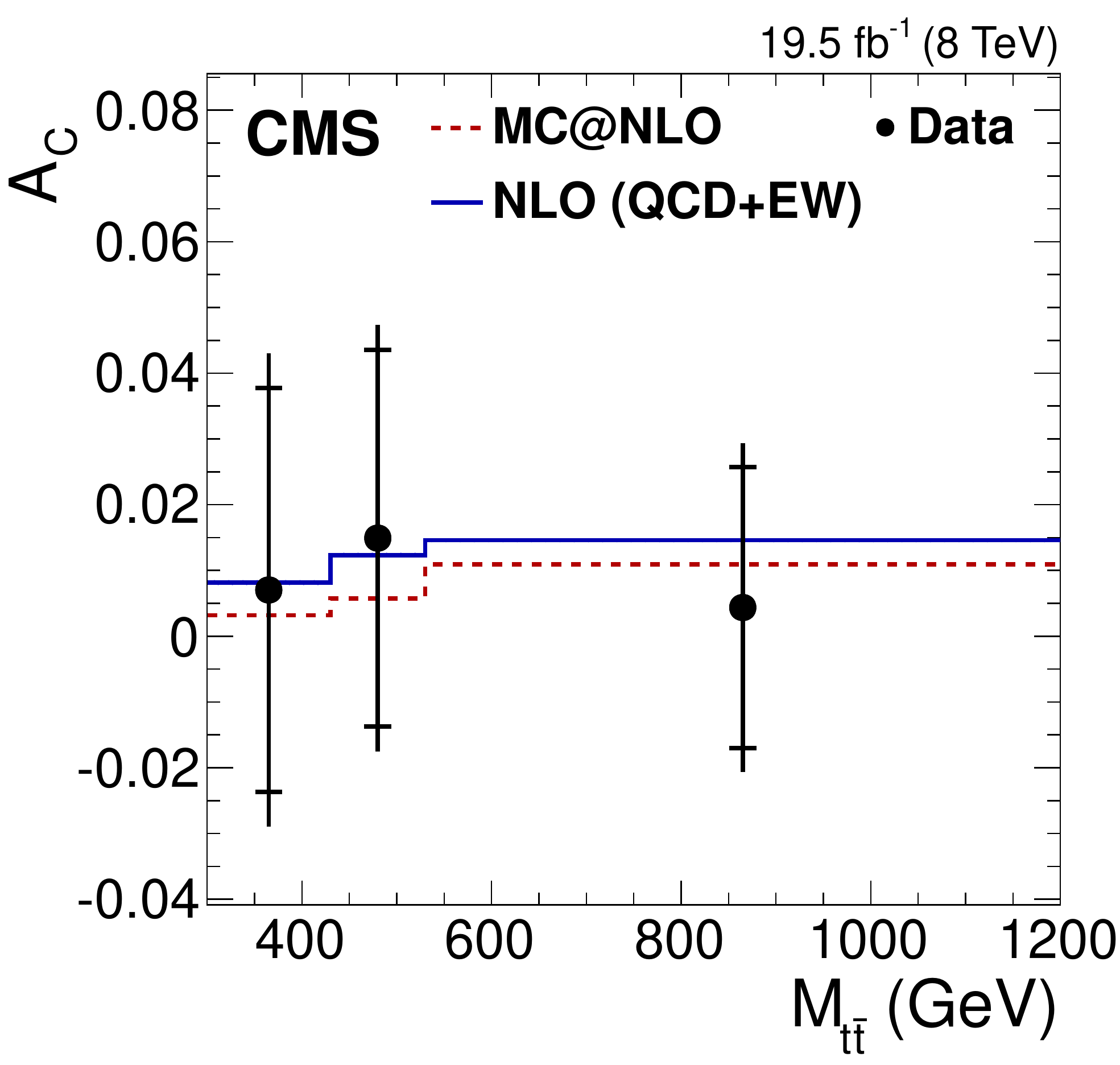}
\includegraphics[width=0.40\textwidth]{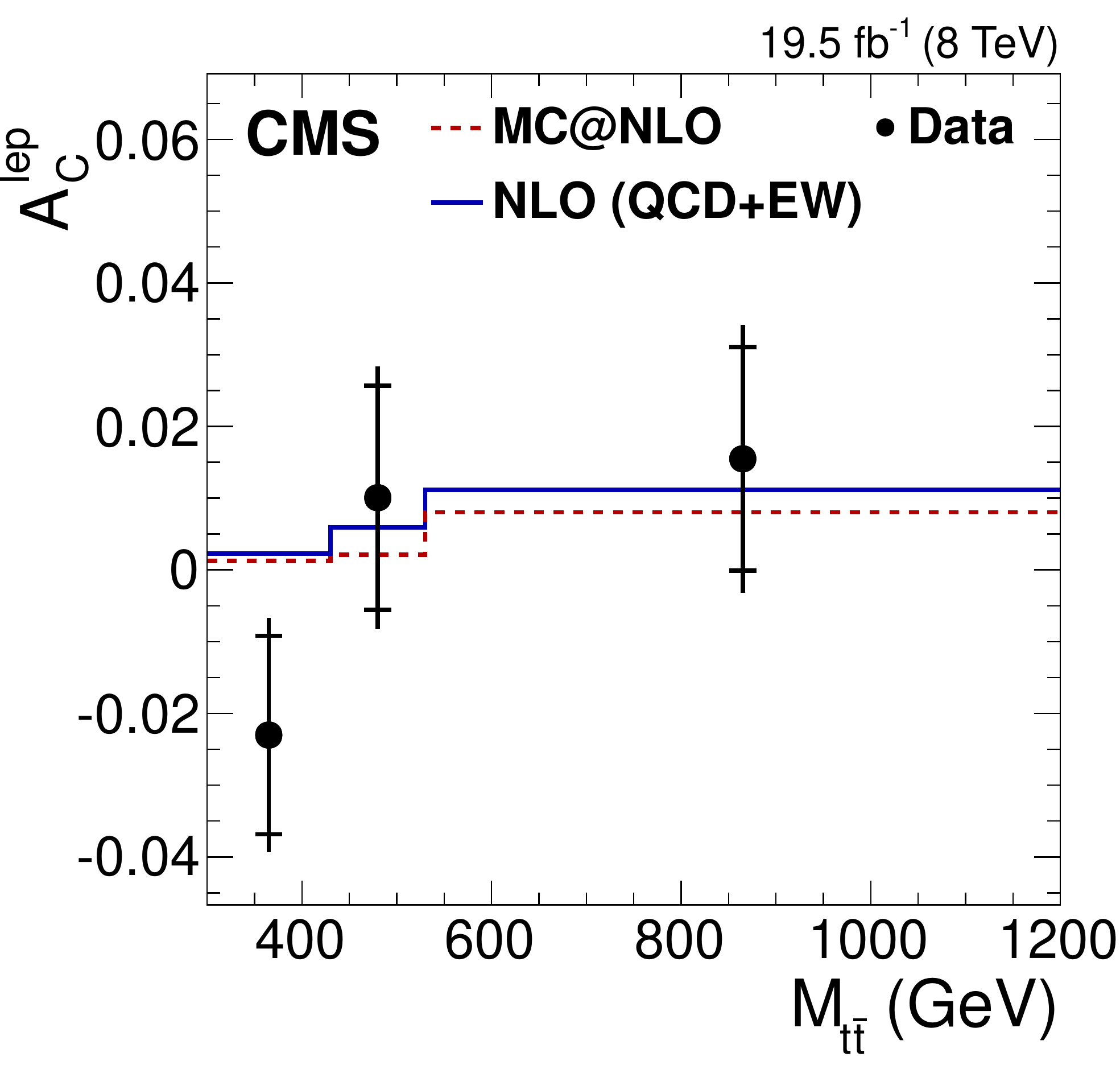}
\includegraphics[width=0.40\textwidth]{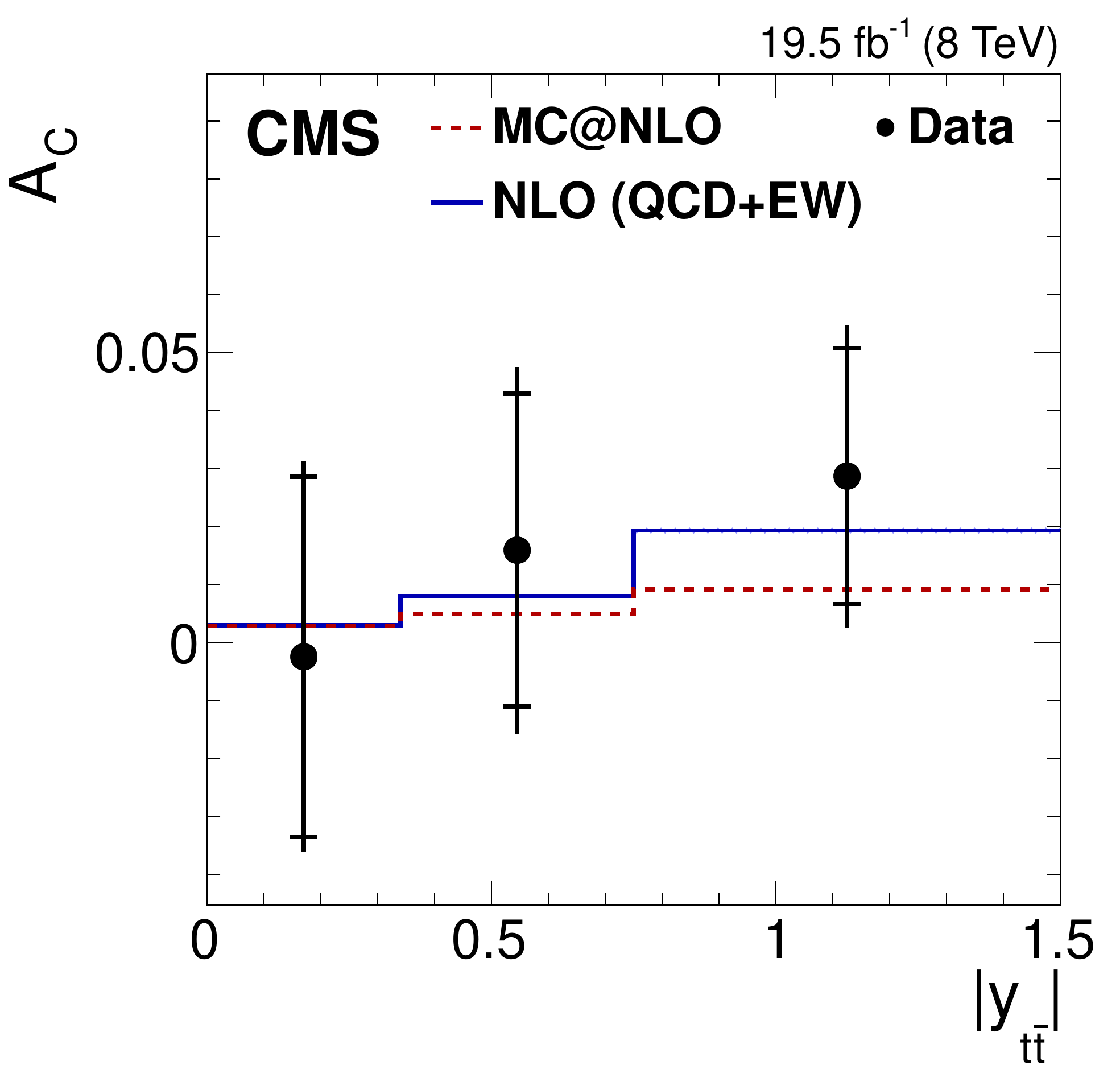}
\includegraphics[width=0.40\textwidth]{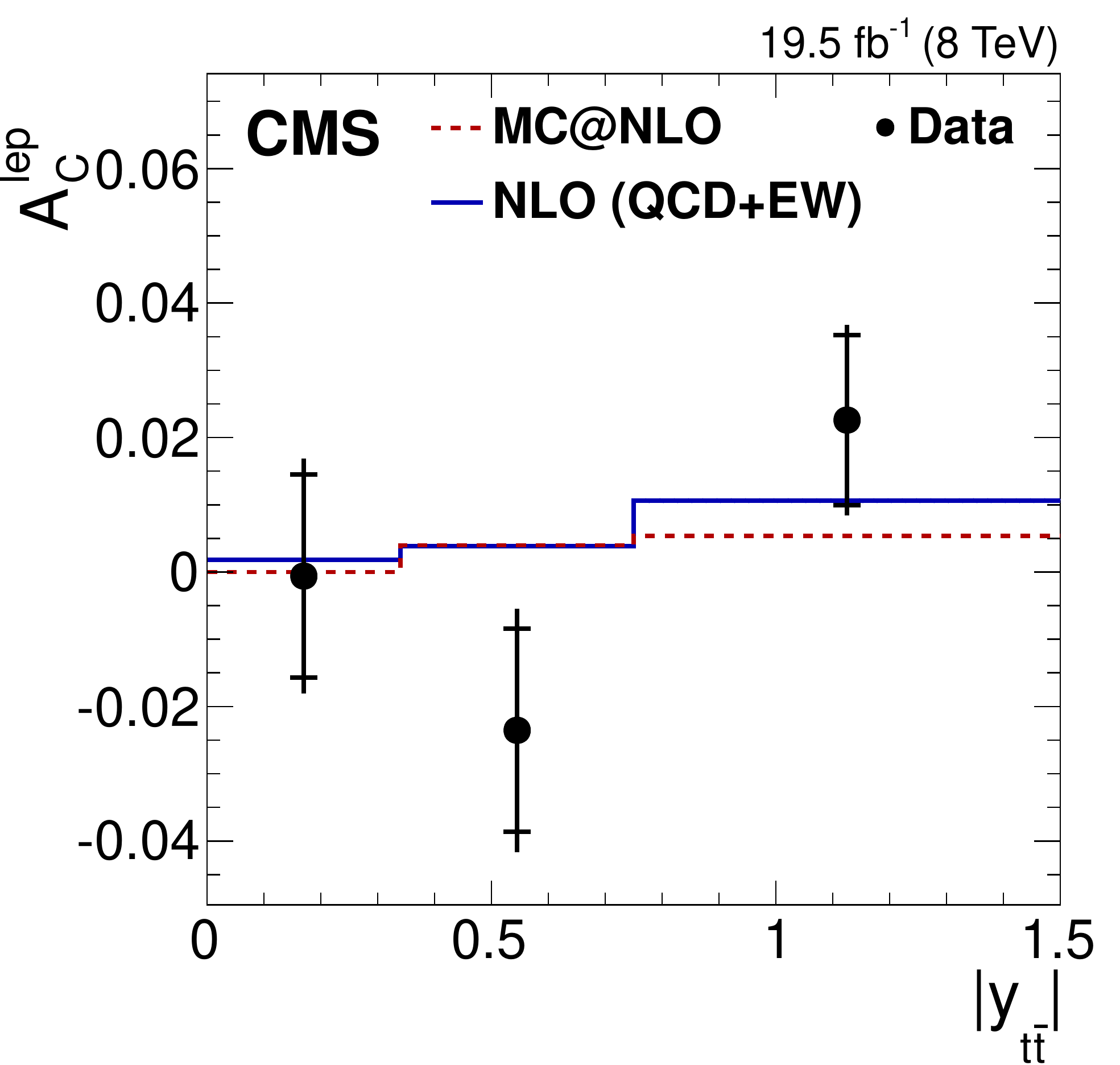}
\includegraphics[width=0.40\textwidth]{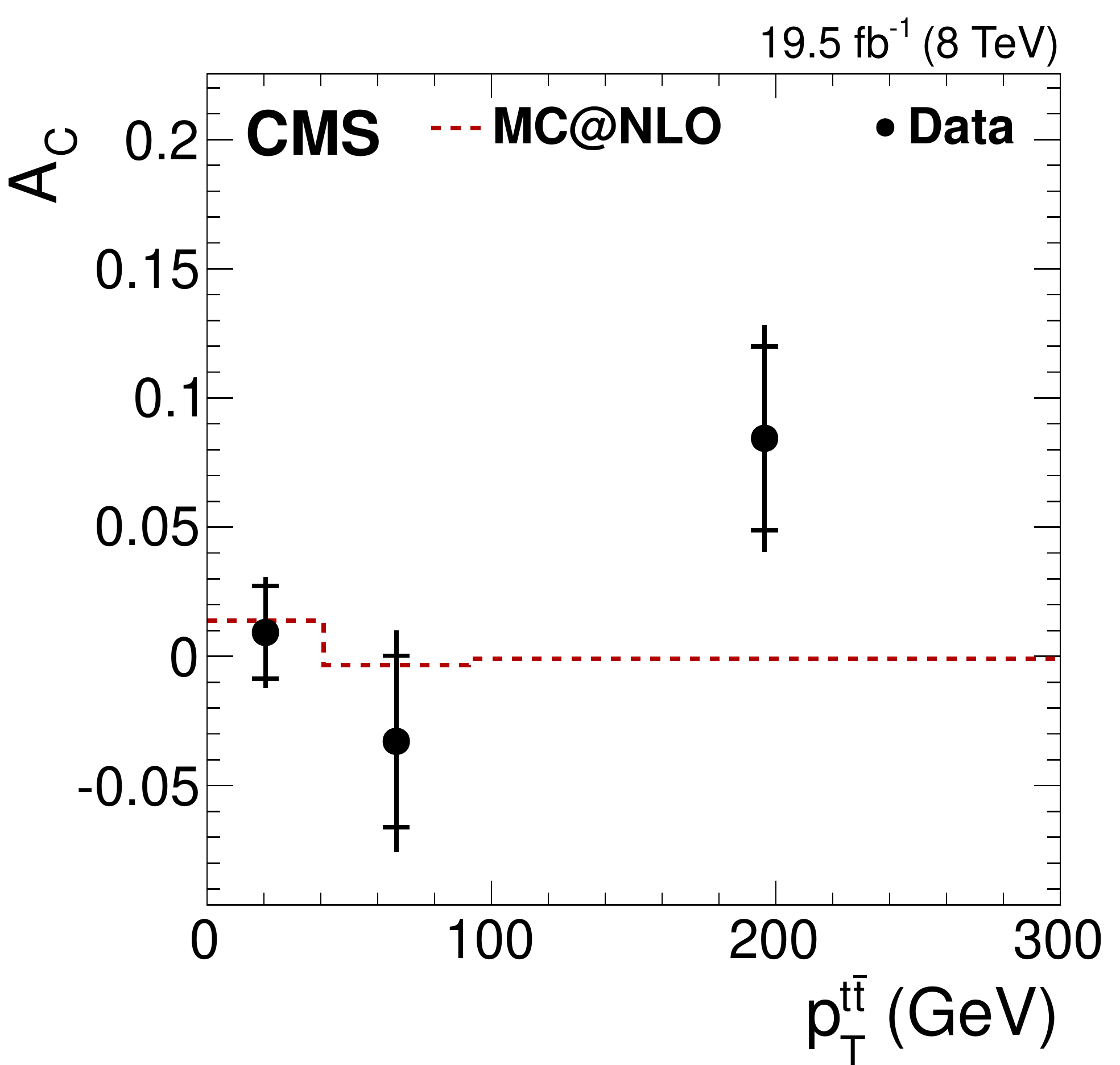}
\includegraphics[width=0.40\textwidth]{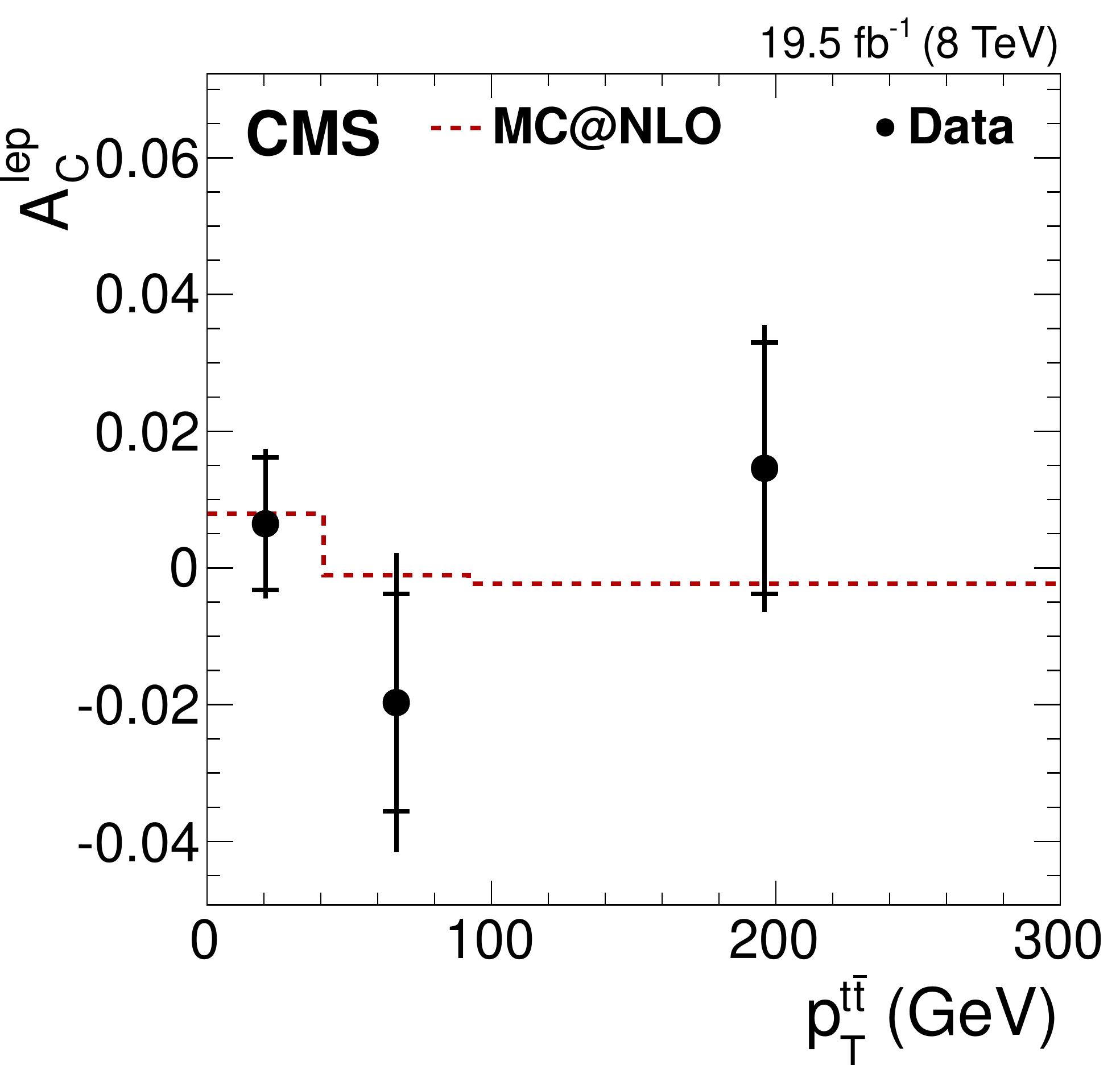}
\caption{\label{fig:AlepCdiff}\protect
Dependence of the \ttbar and leptonic charge asymmetries $A_{\mathrm{C}}$ (left) and $A^{\text{lep}}_{\mathrm{C}}$ (right) obtained from the unfolded distributions in data (points) on $M_{\ttbar}$ (upper), $\abs{y_{\ttbar}}$ (middle), and
$\pt^{\ttbar}$ (lower).
Parton-level predictions from the \MCATNLO simulation and calculations at NLO (QCD+EW)~\cite{ref:BernChargeAsym} are shown by dashed and solid histograms, respectively.
The vertical bars show the total uncertainty, the statistical component of which is marked by a horizontal tick.
The last bin of each plot includes overflow events.
}
\end{figure*}

The unfolded normalized differential cross section from the selected events in data
is shown as a function of $\Delta\abs{y_{\cPqt}}$ and $\Delta\abs{\eta_{\ell}}$ in \fig~\ref{fig:ResultsUnfolded},
along with the parton-level predictions for \ttbar\ production obtained from calculations at NLO in the SM gauge couplings (QCD+EW)~\cite{ref:BernChargeAsym} and with the \MCATNLO\ generator (which does not include EW corrections).
The corresponding $A_{\mathrm{C}}$ and $A^{\text{lep}}_{\mathrm{C}}$ values are presented in \tab~\ref{tab:ResultsUnfolded}.
Correlations between the contents of different bins, introduced by the unfolding process and from the systematic uncertainties, are accounted for in the calculation of the uncertainties.
The measured values are consistent with the expectations from the SM.
The charge asymmetries as a function of $M_{\ttbar}$, $\abs{y_{\ttbar}}$, and $\pt^{\ttbar}$ are also measured. The results, which are shown in \fig~\ref{fig:AlepCdiff},
are consistent with the \MCATNLO\ simulation predictions, as well as with the NLO (QCD+EW) calculations for the $M_{\ttbar}$ and $\abs{y_{\ttbar}}$ dependencies.
No comparison is made with NLO calculations for the $\pt^{\ttbar}$ dependencies as it is expected that the effect of the parton shower
process on the $\pt^{\ttbar}$ distribution makes fixed-order calculations an inadequate approximation of the data.

\section{Summary}

Measurements are presented of the charge asymmetry in \ttbar\ dilepton final states from distributions, unfolded to the parton level, of the absolute rapidity (pseudorapidity) difference of top quarks (leptons) with positive and negative charge.
The data sample corresponds to an integrated luminosity of 19.5\fbinv from pp collisions at $\sqrt{s}=8$\TeV, collected by the CMS
experiment at the LHC.
The \ttbar and leptonic inclusive charge asymmetries are found to be, respectively, $0.011 \pm 0.011\stat \pm 0.007\syst$ and $0.003 \pm 0.006\stat \pm 0.003\syst$ when measured inclusively.
The charge asymmetries are also measured as a function of the invariant mass, absolute rapidity, and transverse momentum
of the \ttbar\ system in the laboratory frame.
Although statistically limited, all measurements are in agreement with the standard model predictions.
Future measurements at $\sqrt{s}=13$\TeV with larger data sets are expected to have better statistical precision outweighing the dilution of the charge asymmetry from the decreased fraction of events with the quark-antiquark initial state.

\begin{acknowledgments}
We would like to thank W.~Bernreuther and Z.-G.~Si for calculating the theoretical distributions shown in this paper.
We congratulate our colleagues in the CERN accelerator departments for the excellent performance of the LHC and thank the technical and administrative staffs at CERN and at other CMS institutes for their contributions to the success of the CMS effort. In addition, we gratefully acknowledge the computing centres and personnel of the Worldwide LHC Computing Grid for delivering so effectively the computing infrastructure essential to our analyses. Finally, we acknowledge the enduring support for the construction and operation of the LHC and the CMS detector provided by the following funding agencies: BMWFW and FWF (Austria); FNRS and FWO (Belgium); CNPq, CAPES, FAPERJ, and FAPESP (Brazil); MES (Bulgaria); CERN; CAS, MoST, and NSFC (China); COLCIENCIAS (Colombia); MSES and CSF (Croatia); RPF (Cyprus); MoER, ERC IUT and ERDF (Estonia); Academy of Finland, MEC, and HIP (Finland); CEA and CNRS/IN2P3 (France); BMBF, DFG, and HGF (Germany); GSRT (Greece); OTKA and NIH (Hungary); DAE and DST (India); IPM (Iran); SFI (Ireland); INFN (Italy); MSIP and NRF (Republic of Korea); LAS (Lithuania); MOE and UM (Malaysia); CINVESTAV, CONACYT, SEP, and UASLP-FAI (Mexico); MBIE (New Zealand); PAEC (Pakistan); MSHE and NSC (Poland); FCT (Portugal); JINR (Dubna); MON, RosAtom, RAS and RFBR (Russia); MESTD (Serbia); SEIDI and CPAN (Spain); Swiss Funding Agencies (Switzerland); MST (Taipei); ThEPCenter, IPST, STAR and NSTDA (Thailand); TUBITAK and TAEK (Turkey); NASU and SFFR (Ukraine); STFC (United Kingdom); DOE and NSF (USA).
Individuals have received support from the Marie-Curie programme and the European Research Council and EPLANET (European Union); the Leventis Foundation; the A. P. Sloan Foundation; the Alexander von Humboldt Foundation; the Belgian Federal Science Policy Office; the Fonds pour la Formation \`a la Recherche dans l'Industrie et dans l'Agriculture (FRIA-Belgium); the Agentschap voor Innovatie door Wetenschap en Technologie (IWT-Belgium); the Ministry of Education, Youth and Sports (MEYS) of the Czech Republic; the Council of Science and Industrial Research, India; the HOMING PLUS programme of the Foundation for Polish Science, cofinanced from European Union, Regional Development Fund; the OPUS programme of the National Science Center (Poland); the Compagnia di San Paolo (Torino); MIUR project 20108T4XTM (Italy); the Thalis and Aristeia programmes cofinanced by EU-ESF and the Greek NSRF; the National Priorities Research Program by Qatar National Research Fund; the Rachadapisek Sompot Fund for Postdoctoral Fellowship, Chulalongkorn University (Thailand); the Chulalongkorn Academic into Its 2nd Century Project Advancement Project (Thailand); and the Welch Foundation, contract C-1845.
\end{acknowledgments}

\clearpage

\bibliography{auto_generated}   

\ifthenelse{\boolean{cms@external}}{}{
\clearpage
\numberwithin{figure}{section}
\appendix
\section[Supplementary material]{Supplementary material\label{app:suppMat}}
\begin{figure*}[!htpb]
\centering
\includegraphics[width=0.49\linewidth]{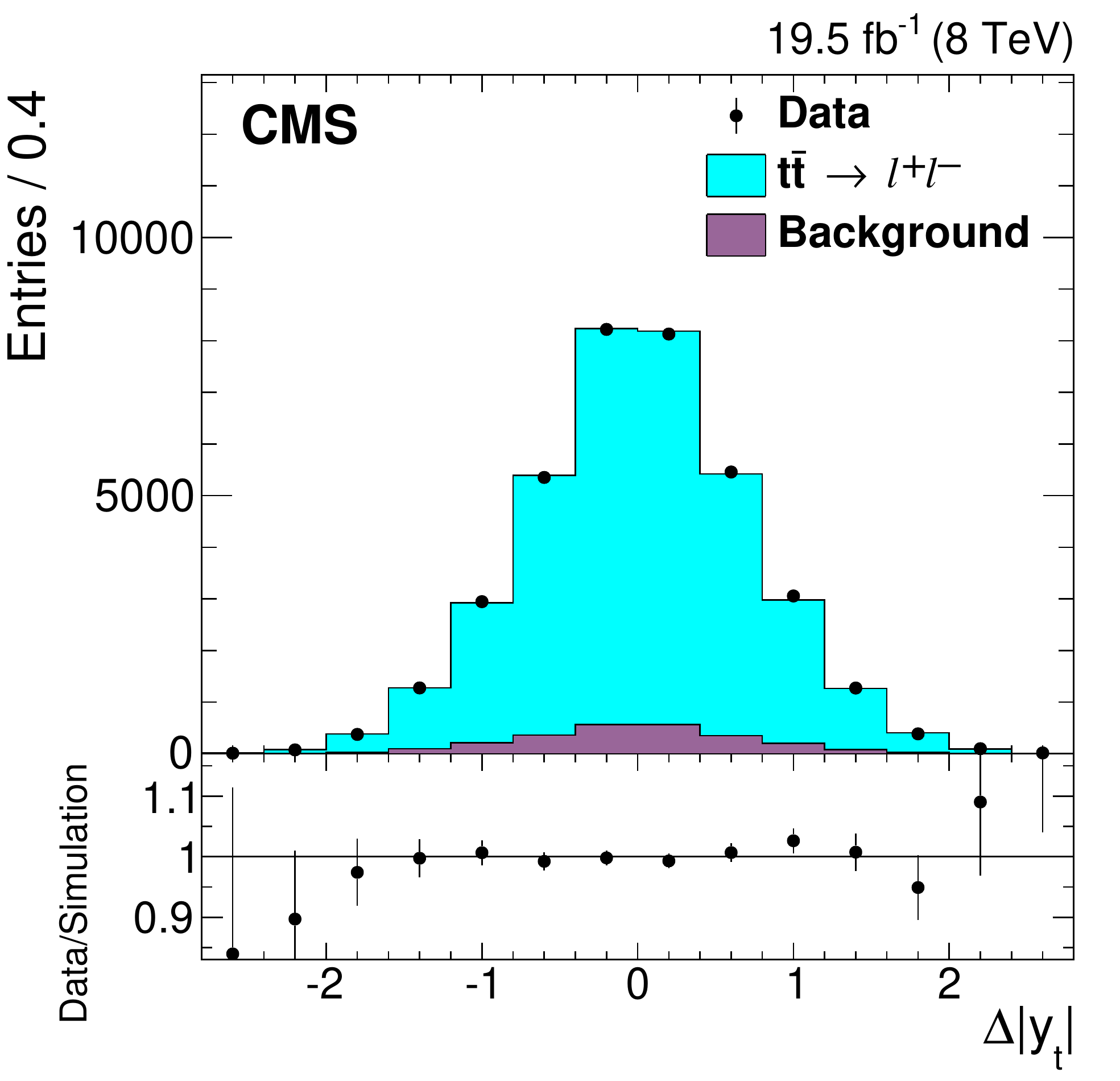}
\includegraphics[width=0.49\linewidth]{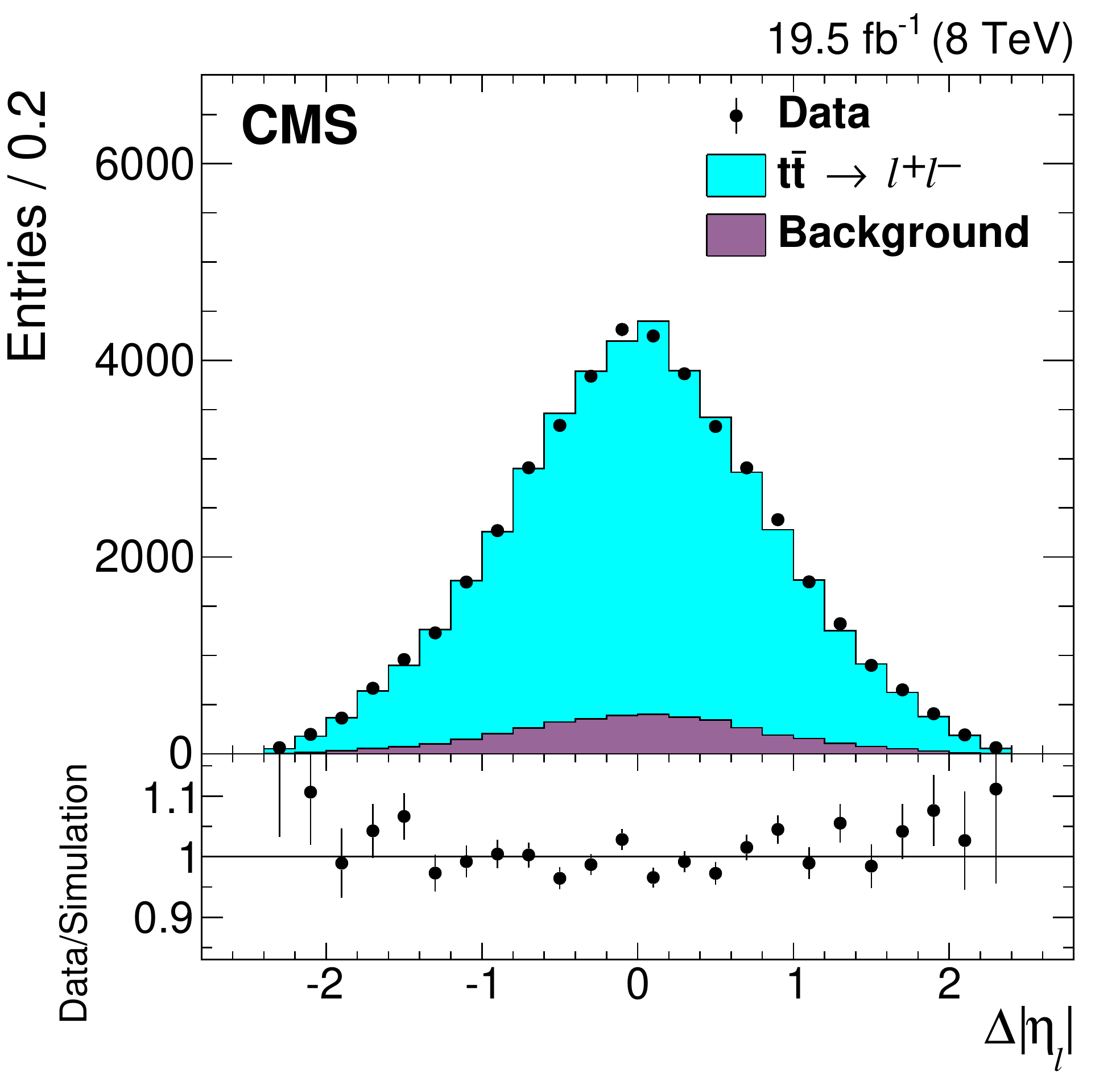}
\caption{
Reconstructed $\Delta\abs{y_{\cPqt}}$ and $\Delta\abs{\eta_{\ell}}$ distributions from data (points) and simulation (histogram), with the predicted signal ($\ttbar\to\ell^{+}\ell^{-}$) and background distributions shown separately.
The simulated signal yield is normalized to that of the background-subtracted data.
The vertical bars on the data points represent the statistical uncertainties.
The lower panels show the ratio of the numbers of events from data and simulation.
}
\end{figure*}
}

\cleardoublepage \section{The CMS Collaboration \label{app:collab}}\begin{sloppypar}\hyphenpenalty=5000\widowpenalty=500\clubpenalty=5000\textbf{Yerevan Physics Institute,  Yerevan,  Armenia}\\*[0pt]
V.~Khachatryan, A.M.~Sirunyan, A.~Tumasyan
\vskip\cmsinstskip
\textbf{Institut f\"{u}r Hochenergiephysik der OeAW,  Wien,  Austria}\\*[0pt]
W.~Adam, E.~Asilar, T.~Bergauer, J.~Brandstetter, E.~Brondolin, M.~Dragicevic, J.~Er\"{o}, M.~Flechl, M.~Friedl, R.~Fr\"{u}hwirth\cmsAuthorMark{1}, V.M.~Ghete, C.~Hartl, N.~H\"{o}rmann, J.~Hrubec, M.~Jeitler\cmsAuthorMark{1}, A.~K\"{o}nig, M.~Krammer\cmsAuthorMark{1}, I.~Kr\"{a}tschmer, D.~Liko, T.~Matsushita, I.~Mikulec, D.~Rabady, N.~Rad, B.~Rahbaran, H.~Rohringer, J.~Schieck\cmsAuthorMark{1}, R.~Sch\"{o}fbeck, J.~Strauss, W.~Treberer-Treberspurg, W.~Waltenberger, C.-E.~Wulz\cmsAuthorMark{1}
% ...
\vskip\cmsinstskip
% ...
1:~~Also at Vienna University of Technology, Vienna, Austria\\
% ...

\end{sloppypar}
\end{document}